 \documentclass[12pt]{revtex4}
\usepackage{graphicx}
\usepackage{amssymb}
\usepackage{amsmath}
\usepackage{amsfonts}
\usepackage{epstopdf}
\usepackage{epsfig}
\usepackage{wrapfig}

\newtheorem{theorem}{Theorem}

\newtheorem{definition}{Definition}

\begin{document}

\title{{\LARGE A BQP-complete problem related to the Ising model partition function via a new connection between quantum circuits and graphs}}
\author{Joseph Geraci}
\affiliation{University of Toronto \\ Department of Mathematics}
\affiliation{University of Southern California \\ CQIST - Center for Quantum Information Science and Technology}
\maketitle

\section*{\protect\underline{{\protect\Large Abstract}}}
We present a simple construction that maps quantum circuits to graphs and vice-versa. Inspired by the results of D.A. Lidar linking the Ising partition function with quadratically signed weight enumerators (QWGTs), we also present a BQP-complete problem for the additive approximation of a function over hypergraphs related to the generating function of Eulerian subgraphs for ordinary graphs. We discuss connections with the Ising partition function.
\newpage

\section{Introduction}

Relationships between quantum computation and graph theory are emerging beyond applications to graph theoretic problems. For example in the one-way quantum computation setting, quantum graph states and their relationship to entanglement is already well known \cite{graph-state}. Here one sees the correspondence between a graph and a quantum state where the vertices correspond to the qubits of the state and the edges correspond to pairs of qubits. For quantum circuits one may make use of a graph to actually represent the circuit or architecture \cite{qc-graph}. For example, if we let $\Gamma=(V,E)$ be a graph, then the set of vertices $V$ may represent the individual qubits (or input into the circuit) and the edges $E$ correspond to any pair of qubits that may be acted upon by a two qubit gate. In another related approach provided in \cite{tensor},they instruct to ``regard each gate as a vertex, and for each input/output wire add a new vertex to the open edge of the wire.'' Using this correspondence they prove statements about families of quantum circuits that are classically simulatable. For the work here we assume the usual quantum circuit model and that one may efficiently swap between different universal gate sets due to the famous theorem of Solovay and Kitaev \cite{Nielsen}. The methods used here have been employed to study quantum circuits which may be simulated classically \cite{JOE2}.

In this paper we provide a simple construction which makes a direct connection between quantum circuits and graphs via their incidence structure. As an application we construct a function related to the generating function of Eulerian subgraphs and discuss the relationship to the Ising partition function. We show that quantum computers can provide additive approximations of this related generating function, which we call the \emph{signed} generating function of Eulerian subgraphs, $E'(\Gamma, \lambda)$. We demonstrate that it is a BQP-complete problem (when we allow it to be defined over hypergraphs which are a generalization of graphs) as it is intimately related to quadratically signed weight enumerators (QWGTs) \cite{Laflamme} via this construction. It is well known that the Ising partition function $Z$ may be expressed in terms of the generating function of Eulerian subgraphs, $E(G,\lambda)$.  We provide some ideas for the future to use $E'(\Gamma, \lambda)$ for efficient additive approximations of $Z$. Recently in \cite{dorit-tutte}, an additive approximation algorithm of the Tutte polynomial was given which solved instances shown to be BQP-complete. As the Ising partition function is just a specialization of the Tutte polynomial, the complexity of additive approximations of certain non-planar instances of the Ising partition function is an interesting open problem.

\subsection{Ising spin model}

We shall first introduce the Ising spin model in order to motivate the computation of generating function of Eulerian subgraphs. Let $G=(E,V)$ be a finite, arbitrary undirected
graph with $|E|$ edges and $|V|$ vertices. In the Ising model, each vertex is associated with a classical spin ($\sigma _{i}=\pm 1$) and
each edge $(i,j)\in E$ with a bond ($J_{ij}=\pm J$). The Hamiltonian of the
spin system is
\begin{equation} H(\sigma )=-\sum_{(i,j)\in E}J_{ij}\sigma _{i}\sigma
_{j}. \label{Ham} \end{equation} 
The probability of the spin configuration $\sigma $ in thermal
equilibrium at temperature $T$ is given by the Gibbs distribution: $P(\sigma
)={\frac{1}{Z}}W(\sigma )$, where the Boltzmann weight is $W(\sigma )=\exp
[-\beta H(\sigma )]$, $\beta =1/kT$, and $Z$ is the partition function: 
\begin{equation}
Z_{\{J_{ij}\}}(\beta )=\sum_{\sigma }\exp [-\beta H(\sigma )].  \label{eq:Z}
\end{equation}

\subsection{Generating function of Eulerian subgraphs}
An Eulerian subgraph of a graph $\Gamma$ is a set of edges that form a tour (a path that begins and ends at the same vertex) in which every vertex is of even degree. The \emph{generating function of Eulerian subgraphs of $\Gamma$} is given by
\[ E(\Gamma, x)=\sum_{a} x^{wt(a)} \] where the sum is over all Eulerian subgraphs and $wt$ is a weight function (in this case the number of edges in the subgraph).   This brings us to another expression for the Ising partition function which was discovered by van der Waerden and is given by
\begin{equation} Z(\beta) = 2^{|V|} \prod_{\{i,j\}\in E} \cosh(\beta J_{ij}) E(\Gamma, \tanh(\beta J_{ij})) \label{genfunc} \end{equation} It is this form of the partition function that motivates this work \cite{eulerian,Welsh}.

\section{QWGTs and their relation to the Ising partition function}\label{sec2} 

\begin{definition} Quadratically Signed Weight Enumerators (QWGTs) in general
are of the form \cite{Laflamme} 
\begin{equation}
S(A,B,x,y)=\sum_{b:Ab=0}(-1)^{bBb}x^{|b|}y^{n-|b|},  \label{eq:S}
\end{equation}%
where $A$ and $B$ are $0,1$-matrices with $B$ of dimension $n\times n$ and $%
A $ of dimension $m\times n$. The variable $b$ in the summand ranges over $%
0,1$-column vectors of dimension $n$. All calculations involving $A,B$ or $b$
are done modulo $2$. 
\end{definition}

Note that the evaluation of a QWGT, given $x$ and $y$ are natural numbers, is in general a $\#$P problem and as it includes evaluations of the weight enumerator of binary linear codes it is in fact $\#$P-complete \cite{Laflamme}.

We shall now review how QWGT's were constructed in \cite{Laflamme} in some detail. Let $\mathbf{G}$ be a quantum circuit and $U(\mathbf{G})$ the corresponding unitary operator. Note that a universal gate set can be achieved by allowing arbitrary rotations about any product of Pauli operators  i.e.

\[ e^{-i \sigma_b \theta/2} = \cos \left(\frac{\theta}{2}\right)I + i \sin\left(\frac{\theta}{2}\right)\sigma_b \] where $\sigma_b = \prod_{i=1}^n \sigma_{b_i}^{(i)}$ such that $\sigma_{00} = I$,$\sigma_{01} = \sigma_X$, $\sigma_{11} = \sigma_Y$ and $\sigma_{10} = \sigma_Z$ \cite{vazi}. This means that $b$ is a binary vector whose length is $2n$, twice the number of qubits, and the superscript $(i)$ represents the qubit which is operated on by the corresponding Pauli matrix.

It is possible to express our unitary operator as a product of real gates and we can do this as follows. Take the product $U(\mathbf{G}) = G_NG_{N-1} \cdots G_1$ where each gate is of the form 
\[ G_k = \frac{1}{\gamma} \left(\alpha \pm i \beta \sigma_{b_k} \right) \] where again $b_k$ is a binary vector of length $2n$ but each $b_k$ can only contain an odd number of $11$'s i.e. each gate can only have an odd number of Pauli $Y$ operators $\sigma_Y$. Also note that $\gamma = \sqrt{\alpha^2 + \beta^2}.$ As a further modification, which will allow us to have simple multiplication rules for our gates, define 

\[ \tilde{\sigma}_{b_k} = (-i)^{|b|_Y}\sigma_{b_k} \] where $|b|_Y$ is the number of $\sigma_Y$'s occurring in $\sigma_{b_k}$. 

We can now write \[ G_k = \frac{1}{\gamma}(\alpha + \beta \tilde{\sigma}_{b_k}). \]

Now define $C$ to be the block diagonal matrix whose blocks consist of

\[ \left( \begin{array}{cc}
0 & 1 \\
0 & 0 \\ \end{array} \right). \]

Then the property that $b_k$ has an odd number of $11$'s is  given by $b^tCb=1$ and so we have the multiplication rule

\begin{equation} 
\tilde{\sigma}_{b_1} \tilde{\sigma}_{b_2} = (-1)^{b_1^t C b_2} \tilde{\sigma}_{b_1 + b_2}
\end{equation} where the addition in the subscript is bit by bit modulo 2. 

Let $H$ be the $(2n \times N)$ matrix whose columns are the $b_k$. $H$ is a polynomial size representation of the quantum circuit where each column represents a gate and every pair of rows represents a qubit. We then have the following expansion. 

\begin{eqnarray} U(\mathbf{G}) &=& \prod_{k=N}^1 G_k \\ &=& \prod_{k=N}^1 \frac{1}{\gamma} (\alpha + \beta \tilde{\sigma}_{b_k}) \\ &=& \frac{1}{\gamma^N} \sum_a (-1)^{a^t \mathrm{lwtr}(H^tCH)a} \alpha ^{|a|} \beta^{N-|a|} \tilde{\sigma}_{H_a}
\end{eqnarray}

Now note that if we only sum over the $a$'s such that $CHa=0$ then we assure that
\[ \langle 00 \cdots \lvert U(\mathbf{G}) \rvert 00 \cdots \rangle = \frac{1}{\gamma^N} \sum_a (-1)^{a^t \mathrm{lwtr}(H^tCH)a} \alpha ^{|a|} \beta^{N-|a|} \] is always non-zero for this omits $X$ and $Y$ gates from our sum. 

As a simple example to illustrate the correspondence between the matrix representation $H$ of the circuit and the actual operation of the circuit, consider  

\[ H = \left [ \begin{array}{ccc} 
 			1 & 1 & 1 \\
			0 & 0 & 1 \\
			0 & 1 & 1 \\
			1 & 0 & 0 \\
			1 & 1 & 1 \\
			1 & 1 & 0  \end{array} \right ]\] This matrix represents a circuit which operates in the following way:
\[e^{-iZ^{(1)}\otimes X^{(2)}\otimes Y^{(3)}\frac{\theta}{2}}e^{-iZ^{(1)}\otimes Z^{(2)}\otimes Y^{(3)}\frac{\theta}{2}}e^{-iY^{(1)}\otimes Z^{(2)} \otimes Z^{(3)}\frac{\theta}{2}} \] where the superscripts represent which qubit is being acted upon.  Thus each column encodes each exponentiated operator, i.e., each gate. When using our proposed gate set, we would have

\begin{equation*}
\frac{1}{\gamma^3}[(\alpha I -i \beta Z^{(1)}\otimes X^{(2)}\otimes Y^{(3)})(\alpha I -i \beta Z^{(1)}\otimes Z^{(2)}\otimes Y^{(3)})(\alpha I -i\beta Y^{(1)}\otimes Z^{(2)} \otimes Z^{(3)})] 
\end{equation*}

In \cite{Lidar2} it was shown that the Ising partition function can be expressed in terms of a QWGT. 
Let $A$ be the incidence matrix of a graph $g$, i.e.,
\begin{equation}
A_{v,(i,j)}=\left\{ 
\begin{array}{ll}
1 & \mbox{$(v=i \:\: {\rm and}\:\: (i,j)\in E)$} \\ 
0 & \mbox{${\rm else}$}%
\end{array}%
\right. .
\end{equation}%

Then we have
\begin{eqnarray} Z_w(\lambda) &=& \frac{2^{|V|}}{(1-\lambda^2)^{|E|/2}} \sum_a (-1)^{a^t B a} \lambda^{|a|}=\frac{2^{|V|}}{(1-\lambda^2)^{|E|/2}}\sum_{a\in \ker A} (-1)^{a\cdot w} \lambda^{|a|} \\
&=&  \frac{2^{|V|}}{(1-\lambda^2)^{|E|/2}} S(A,dg(w),\lambda,1) \label{Z} \end{eqnarray}
where $w=(w_{12},w_{13},\dots)$ ($w$ gives the distribution of ferromagnetic ($w_{ij}=0$) or anti-ferromagnetic ($w_{ij}=1$) interactions along the edges of the given graph), $\lambda = \tanh(\beta J)$ (the ``temperature''), $V$ is the set of vertices, $E$ is the set of edges and $B=dg(w)$ is the diagonal matrix formed by putting $w$ on the diagonal and zeros everywhere else. This form of Z will be considered in future work.

\section{A relationship between hypergraphs and quantum circuits via QWGTs}
The following ``mapping'' between hypergraphs and quantum circuits was first introduced in order to find a way to compute the Ising partition function. It was extended to obtain a class of quantum circuits which can be simulated classically in \cite{JOE2}  and is based on equation (\ref{Z}). This mapping involves interpreting the matrix representation $H$ of the quantum circuit, as outlined above, as coming from the incidence matrix of a hypergraph. In this way we have a many to one mapping from quantum circuits to a hypergraph. 

First we define hypergraphs. 

\begin{definition}

A hypergraph is a generalization of a graph where edges are replaced by \emph{hyperedges}. Let $V=\{v_1,v_2, \dots, v_k\}$ be the set of vertices and let $E=\{e_1,e_2, \dots, e_n\}$ be the set of hyperedges. Each $e_i=\{v_{i1},v_{i2}, \dots, v_{im} \}$ is a collection of vertices where each $v_{ij} \in V$. 
\end{definition}

The standard reference for hypergraphs is \cite{berge:book}. Note that graphs are just a special case of hypergraphs where each edge just consists of two vertices and that the incidence matrix is defined in the same manner as above. 

For the description of the mapping in the next subsection, we shall restrict our quantum circuits so that the corresponding hypergraphs are ordinary graphs. We shall point out when this restriction is important.

\subsection{ THE MAPPING}

The motivation for this mapping is to obtain a QWGT equal to a matrix element of the unitary matrix of a quantum circuit that looks something like the generating function of Eulerian subgraphs $E(\Gamma,\lambda)$. If we were able to efficiently approximate $E(\Gamma,\lambda)$ then according to equation (\ref{genfunc}) we would have an efficient method of approximating the Ising partition function.

It turns out that if we take the ansatz  \begin{equation} G_k = \frac{1}{\sqrt{\lambda^2 + 1}}(\lambda + \tilde{\sigma}_{b_k}) \label{ansatz} \end{equation}
for the gate set we obtain   

\begin{eqnarray} U(\mathbf{G}) &=& \prod_{k=N}^1 \frac{1}{(\lambda^2 + 1)}(\lambda + \tilde{\sigma}_{b_k}) \\ &=&  \frac{1}{(\lambda^2 + 1)} \prod_{k=N}^1(\lambda + \tilde{\sigma}_{b_k}) \\ &=& \frac{1}{(\lambda^2 + 1)^{N/2}} \sum_a (-1)^{a^t \mathrm{lwtr}(H^tCH)a} \lambda^{|a|} \tilde{\sigma}_{H_a}. \label{U} \end{eqnarray}  Now, ignoring the normalization we have  

\[\langle 00 \cdots \lvert U(\mathbf{G}) \rvert 00 \cdots \rangle \propto  \sum_{a\in \mathrm{Ker}(CH)} (-1)^{a^t \mathrm{lwtr}(H^tCH)a} \lambda^{|a|} \] 

Let us make the following assumptions:

\begin{enumerate} 
\item Take $CH$ to be a binary matrix with only two 1's per column i.e. $CH$ will be identified with the incidence matrix of the given graph $\Gamma$. \emph{For hypergraphs, this is not necessary as the columns of a hypergraph incidence matrix may be populated by more than two 1's, as this corresponds to edges consisting of multiple vertices.}
\item $H$ is a matrix of dimension $2n \times N$ with one (11) and at most one (01) per column, i.e., one $Y$ operation and  at most one $X$ operation per gate respectively. The source of this restriction will be explained below. For example a column may look like 
\[ \left (\begin{array}{c}
1  1  0   0  0  1  1  0  0  0  1  0 \end{array} \right )^T.
\]  H encodes the quantum circuit. \emph{For hypergraphs, these restriction vanish, however it is necessary that there are an odd number of (11)'s per column as our gate set depends on this restriction.}

These assumptions will provide the basis for a natural mapping between quantum circuits and graphs. Consider one additional assumption.
 
\item 
\begin{equation}
a^{t}\mathrm{lwtr}(H^{t}CH)a=0 \mod 2~~\forall a\in \ker
(CH).  \label{euler-condition}
\end{equation}
\end{enumerate}

This ensures that the matrix element $\langle 00 \cdots \lvert U(\mathbf{G})\rvert 00 \cdots \rangle$ is equal to the generating function of eulerian subgraphs. This is achieved as follows. 

First we need to associate the incidence matrix $A$ with a matrix $CH$. The only thing we need to do is to create a matrix with double the number of rows of $A$, with row $2i-1$ occupied by the $i^{th}$ row of $A$, and each even row the zero vector. Thus we obtain 
\[ CH = \left( \begin{array}{ccccc}
A_{11} & A_{12} & \dots & A_{1N}\\
0 & 0 & \dots & 0 \\
A_{21} & A_{22} & \dots & A_{2N} \\
\vdots & \vdots & \dots & \vdots\\ 0 & 0 & \dots & 0
\end{array} \right)\] As far as the graph is concerned, this amounts to adding isolated vertices, which does not add any cycles. We now have the $2n \times N$ matrix $CH$ as our representation for $\Gamma$. 
 By the action of $C$, we see that $CH$ gives us some freedom in our choice of $H$, which is the matrix representation of the quantum circuit. Specifically we have
\[ H = \left( \begin{array}{ccccc}
x_1 & x_2 & \dots & x_N \\
A_{11} & A_{12} & \dots & A_{1N}\\
x_{N+1} & x_{N+2} & \dots & x_{2N} \\
A_{21} & A_{22} & \dots & A_{2N} \\
\vdots & \vdots & \dots & \vdots\\ 
A_{n1} & A_{n2} & \dots & A_{nn}  
\end{array} \right)\] The $x_i$ must be selected according to the constraints mentioned above. We see that column $k$ of $H$ will only have two $A_{ik}{'s}$ which are equal to 1 as these come from an incidence matrix and by definition, an incidence matrix has only two 1's per column. One 1 is possible as this represents a loop, i.e., an edge that begins and terminates at the same vertex. Further, by the QWGT formalism constructed above, we must have an odd number of $Y$ ((11) entry in the column) operations per gate. Hence, there must be one 11 per column in the matrix $H$. There are only two positions where we could select an $x_i$ in column $k$ to be 1 such that it will be followed by an $A_{jk}$ that is equal to 1.  Thus each column (or gate) must have only one $Y$ operation. By the same reasoning we see that there is only one possible place to put an $X$ operation, i.e., only one way to place a (01) in column $k$. So there can be at most one $X$ operation per gate. There is no restriction as to the number of $Z$ operations per gate as we have the freedom of putting a 1 before any $A_{ik}$ that is set to 0. By turning the $x_i$ on or off (1 or 0 respectively) we obtain different circuits. This provides a degree of freedom that allows one to choose a quantum circuit that may satisfy the final assumption which ensures that the sum $\sum_{a\in \mathrm{Ker}(CH)} (-1)^{a^t \mathrm{lwtr}(H^tCH)a} \lambda^{|a|}$ is equal to $\sum_{a\in \mathrm{Ker}(CH)} \lambda^{|a|}$ which is $E(\Gamma,\lambda)$ as desired. 

Without the restriction given by 
\[
a^{t}\mathrm{lwtr}(H^{t}CH)a=0 \mod 2~~\forall a\in \ker
(CH),
\]
we actually have that \[ \langle 00 \cdots \lvert U(\mathbf{G}) \rvert 00 \cdots \rangle = \frac{1}{(\lambda^2 + 1)^{|E|/2}} \sum_{a\in \mathrm{Ker}(CH)} (-1)^{h_a}\lambda^{|a|}. \] This means that knowledge of the matrix element $\langle 00 \cdots \lvert U(\mathbf{G}) \rvert 00 \cdots \rangle$ amounts to knowledge of \begin{equation} E'(\Gamma,\lambda) = \sum_{a\in \mathrm{Ker}(CH)} (-1)^{h_a}\lambda^{|a|} \end{equation} where the $h_a$ refer to the $a^{t}\mathrm{lwtr}(H^{t}CH)a$. We shall call $E'(\Gamma,\lambda)$ the \emph{signed} generating function of Eulerian subgraphs as the sum is over all subgraphs whose edges have even degree. Specifically, as $CH$ is associated with the incidence matrix of a graph, the whole null space of $CH$ are the characteristic vectors of all Eulerian subgraphs\cite{pfaffian}.

\section{BQP-completeness}
\begin{theorem} Additive approximations of $E'(\Gamma, \lambda)$ over hypergraphs is BQP-complete.
\end{theorem}

Before proving this theorem we would like to note that we include evaluations of $E'(\Gamma,\lambda)$ for hypergraphs $\Gamma$ only for completion, but that this is not necessary. The restriction to graphs forces the corresponding gates to allow an $X$ operation on one qubit, and forces one to have a $Y$ operation on another qubit
, but an arbitrary number of $Z$ operations on the remaining qubits. As these gates correspond to exponentiated Pauli operators, these are multi-qubit operations and thus it is easy to implement entanglement under this restriction as well as control gates. Thus, from the results in \cite{gates1,gates2} we see that the quantum circuits corresponding to ordinary graphs are capable of universal quantum computation. In addition, as our mapping depends on the sum over all simple cycles of a given graph, any one qubit operation may be inserted without effecting the sum, as these correspond to adding loops, i.e., an edge that begins and ends at the same vertex.

{\bf{Proof of Theorem 1:}}

It is clear from the above that an approximation of the matrix element $\langle 00 \cdots \lvert U(\mathbf{G}) \rvert 00 \cdots \rangle$ will give an approximation to $E'(\Gamma,\lambda)$. Recall from \cite{dorit-bqp,jordan} that via the Hadamard test one can obtain an additive approximation of this matrix element. This means that using a quantum computer we can retrieve an approximation $m$ (with some probability of success bounded below by .75, say) which satisfies 
\[ \langle 00 \cdots \lvert U(\mathbf{G}) \rvert 00 \cdots \rangle - p < m < \langle 00 \cdots \lvert U(\mathbf{G}) \rvert 00 \cdots \rangle + p \] where $p$ is a polynomially small parameter. (This intuitive form can be easily derived via the definition of an additive approximation provided in \cite{counting}.) 

Now, assuming that we have at our disposal the universal gate set given by $\theta = \pm 2 \arccos(4/5)$ rotations of products of Pauli operators, then with an overhead of $\mathrm{polylog}(|E|/\epsilon)$ of gates we may approximate our gates $G_k$, to accuracy $O(\epsilon/|E|)$ \cite{Laflamme}. This means that we may indeed approximate the signed generator function of Eulerian subgraphs via the Hadamard test and so this problem is in BQP. 

Now, the other direction. We must demonstrate that knowledge of $E'(\Gamma,\lambda)$ is enough to simulate any quantum circuit. First, any quantum circuit corresponds to a hypergraph under the scheme presented above. Since BQP is a decision class all we have to do is convert a quantum circuit into its decision making counterpart. This may be done as follows \cite{z2}.

\begin{figure}[htp]
\centering
\includegraphics[scale=0.5]{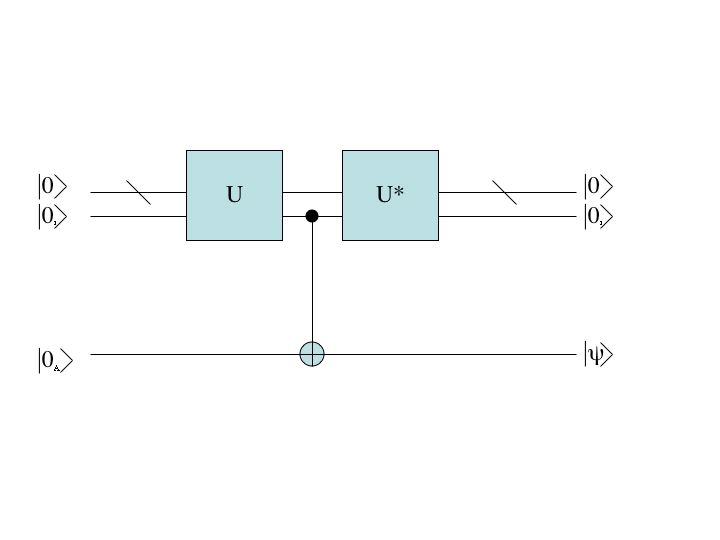}
\vspace{-20pt}
\caption{A circuit illustrating the procedure to apply a quantum circuit $U$ to a decision problem.}\label{circuit}
\end{figure}

Let $U$ be a quantum circuit for some decision problem and without loss of generality, assume that the ``yes'' or ``no'' answer is given by the output of the first qubit $U_1$, i.e., $|0\rangle$ or $|1\rangle$ is ``yes'' or ``no'' respectively. The remaining qubits are ignored and assumed to be extraneous. Now, take an ancilla qubit $U_A$ set to $|0\rangle$ and adjoin it to $U$. Next, CNOT the output of $U_1$ with $U_A$ and encode the answer as the state $|\psi\rangle$. Apply the inverse of the circuit, $U^\dagger$, to all the output qubits except $|\psi\rangle$, to uncompute the outputs of $U$ to $|00\cdots0\rangle$. In this way, one arrives at the state $|00\cdots0\rangle|\psi\rangle$ which will either be $|00\cdots0\rangle|0\rangle$ or $|00\cdots0\rangle|1\rangle$, effectively ``deciding'' the decision problem. Thus one can assume, with no loss of generality, that any quantum circuit that solves some decision problem either outputs $|00\cdots0\rangle|0\rangle$ or $|00\cdots0\rangle|1\rangle$ (see FIG. \ref{circuit}). This argument depends on the quantum circuit being able to output the correct answer with certainty. This is of no matter as a similar argument can be made for a circuit which outputs the correct answer with some constant probability above a half \cite{vazi2}. Thus, knowledge that $\langle 00\cdots \lvert U \rvert 00\cdots \rangle = 1$ implies that the control qubit will be $\lvert 0 \rangle$ and thus $\lvert \psi \rangle = \lvert 0 \rangle$. 

This means that knowledge of $E'(\Gamma,\lambda)$ can be used to effectively decide the decision problem for any quantum circuit as it is proportional to the matrix element $\langle 00\cdots \lvert U \rvert 00\cdots \rangle$. In our case, a natural decision problem would be to decide if $E'(\Gamma, \lambda)$ is bounded above by some constant or to decide its sign. The result of this decision would correspond to $\vert \psi \rangle$ either being $\lvert 0 \rangle$ or $\lvert 1 \rangle $ in FIG. \ref{circuit}  $\blacksquare$

{\bf{Important Caveat:}} Upon careful inspection of the above argument one may find something amiss. We are referring to the idea that perhaps knowledge of $E'(\Gamma,\lambda)$ may in fact not be enough to solve all decision problems in BQP, since only one $\lambda$ is specified. Recall that our universal gate set consists of rotations about products of Pauli operators, with two angles at our disposal, namely $\pm 2 \arccos(4/5)$. When we use the gate set given by the ansatz (\ref{ansatz}), the temperature $\lambda$ plays the role of the angle, and so it seems that there is only one angle available for the gate set $G_k$. In fact, gates of the form (\ref{ansatz}), consist of multiple angles \cite{Laflamme}. A quick calculation verifies the following claims.

\begin{enumerate}
\item If one choses $\lambda = \frac{4}{3}$, then one indeed recovers the rotational angles $\theta = \pm 2\arccos(4/5)$. However, in the scheme outlined in this paper this is not acceptable when one moves to the Ising model, as this particular choice of $\lambda$ means that the \emph{physical} quantity $\beta J$ become complex. In fact one would need $\beta J = \frac{\log 6 + i \pi}{2}$. This is of no matter for the quantity $E'(\Gamma,\lambda)$ but is not acceptable when considering its interpretation as a partition function. However, this does mean that the gate set given by our ansatz is capable of universal quantum computation \cite{Laflamme}.

\item If one chooses, for example, $\lambda = \frac{3}{4}$ then one has access to the two rotational angles $\theta = \pm 2 \arcsin (4/5)$ and thus, we may indeed claim universality for our gate set given by equation (\ref{ansatz}). Further, all quantities are now physically acceptable. 
\end{enumerate}

\subsection{Examples} \label{examples}
Here are two simple yet instructive examples.

1) Let the given quantum circuit be encoded by

\[ H = \left( \begin{array}{cccccc}
1 & 0 & 0 & 0 & 0 & 0 \\
1 & 0 & 0 & 1 & 0 & 0 \\
0 & 1 & 0 & 0 & 0 & 0 \\
0 & 1 & 0 & 0 & 1 & 0 \\
0 & 0 & 1 & 1 & 1 & 1 \\
0 & 0 & 1 & 0 & 1 & 1 \\
0 & 0 & 0 & 1 & 0 & 0 \\
1 & 1 & 1 & 1 & 0 & 0 \end{array} \right ) \]

The incidence matrix (ignoring isolated vertices) can be retrieved easily from $H$ and it is given by

\begin{figure}[htp]
\centering
\includegraphics[scale=0.4]{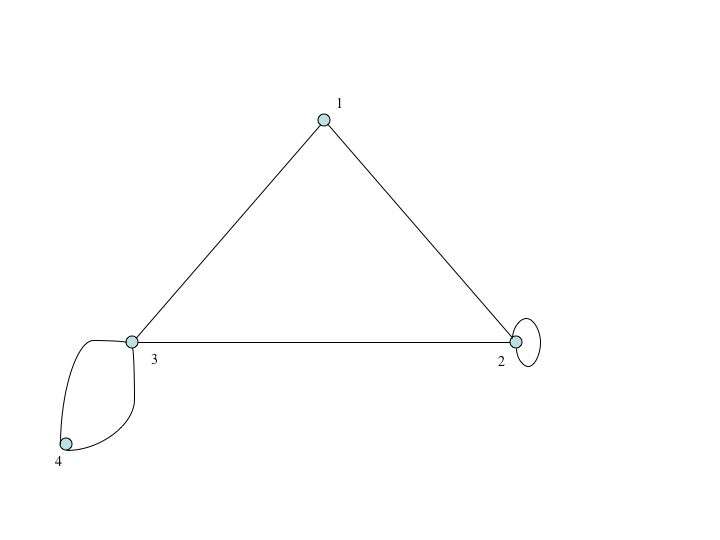}
\caption{The graph obtained from the simple circuit given by H.}\label{decision}
\end{figure}

 \[ \left( \begin{array}{cccccc}
																								1 & 0 & 0 & 1 & 0 & 0 \\
																								0 & 1 & 0 & 0 & 1 & 0 \\
																								0 & 0 & 1 & 0 & 1 & 1 \\
																								1 & 1 & 1 & 1 & 0 & 0 \end{array} \right) \] 
																								
The corresponding graph is given in FIG. \ref{decision}.

2) A simple demonstration of a controlled 2-qubit gate is given by the sign flip operator \cite{Laflamme2} given by 

\[ e^{-i \sigma_z^1 \otimes \sigma_z^2 \frac{\pi}{4}} \] which acts  in the following way

\[ \lvert 0b \rangle \longrightarrow e^{-i\sigma_z^2 \frac{\pi}{4}} \lvert 0b \rangle \] and 
\[ \lvert 1b \rangle \longrightarrow e^{i\sigma_z^2 \frac{\pi}{4}} \lvert 1b \rangle. \] 

In our gate set we would have the corresponding gate, $\alpha I - i \beta \sigma_z^1 \otimes \sigma_z^2$ with the action 

\[ \lvert 0b \rangle \longrightarrow (\alpha - i \beta) \lvert 0b \rangle \] and 
\[ \lvert 1b \rangle \longrightarrow (\alpha + i \beta) \lvert 1b \rangle. \]

\section{Future work: Approximating the Ising partition function}

Recall the following form of the Ising partition function 
\[ Z(\beta) = 2^{|V|} \prod_{\{i,j\}\in E} \cosh(\beta J_{ij}) E(\Gamma, \tanh(\beta J_{ij})) \] and note the difference between the function we are able to approximate via quantum computation, $E'(\Gamma, \lambda)$, and the actual generating function of Eulerian subgraphs $E(\Gamma, \lambda)$. We have that 

\[ E'(\Gamma, \lambda) = \sum_{a\in \mathrm{Ker}(CH)} (-1)^{a^t \mathrm{lwtr}(H^tCH)a} \lambda^{|a|}. \] If we wanted to use this for the approximation of the Ising partition function, as previously mentioned we would require that $a^{t}\mathrm{lwtr}(H^{t}CH)a=0 \mod 2~~\forall a\in \ker(CH)$. If this requirement was met then we would run the quantum approximation algorithm with gate sets corresponding to $\lambda = \tanh(\beta J_{ij}))$. This would require $O(|E|)$ different approximations as indicated by the product over all edges in the above formula. In effect we would have a polynomial additive approximation of the Ising partition function for any set of edge interactions and for any graph. 

But alas, there is a problem. The equation that must be solved (equation (\ref{euler-condition})) in order to ensure that $E'(\Gamma, \lambda) = E(\Gamma, \lambda)$ in fact determines which particular quantum circuit must be used for the computation. By brute force this could require an exponential number of calculations in the number of vertices. Future work will involve studying this approach to see if one can in fact a priori guarantee that equation (\ref{euler-condition}) is satisfied for certain non-planar graphs. For example, if one has knowledge about the parity of all the Eulerian subgraphs (number of edges) then this may be used to efficiently find the representation $H$ of the quantum circuit required. The cubic lattice is an example where every Eulerian subgraph contains an even number of edges. This issue also arises in a very similar approach outlined in \cite{JOE2} but which deals explicitly with equation (\ref{Z}). Other applications of $E'(\Gamma,\lambda)$ will be explored as well as an extension to a two variable function. We will also attempt to use the methods here to find the instances of the Ising model for which evaluations of the partition function is BQP-complete.

\section{Conclusion}

We provide a new way of relating quantum circuits to graphs and vice-versa via an incidence structure of the circuit or graph. We also provide a generating function related to the generating function of Eulerian subgraphs and demonstrate that additive approximations of it for hypergraphs are BQP-complete. Connections to the Ising spin glass partition function were made and a discussion of future work dealing with additive approximations of the Ising partition function was provided.

\begin{acknowledgments}
This work was carried out while under the support of ARO grant
W911NF-05-1-0440 (to D.A. Lidar). I would also like to thank D.A. Lidar for providing me with the opportunity to learn from him. Thanks to Ravi Minhas, Stephan Jordan and Dave Bacon for helpful conversations as well. 
\end{acknowledgments}

\newpage


\end{document}